# Pointwise visual field estimation from optical coherence tomography in glaucoma: a structure-function analysis using deep learning

Authors:
Ruben Hemelings[a,i]*, MS
Bart Elen[i], MS
João Barbosa Breda[a,c,d], MD PhD
Erwin Bellon[f], PhD professor
Matthew B. Blaschko[e], PhD professor
Patrick De Boever[g,h,i], PhD professor
Ingeborg Stalmans[a,b], MD PhD professor

Affiliations:
[a] Research Group Ophthalmology, Department of Neurosciences, KU Leuven, Herestraat 49, 3000 Leuven, Belgium
[b] Ophthalmology Department, UZ Leuven, Herestraat 49, 3000 Leuven, Belgium
[c] Cardiovascular R&D Center, Faculty of Medicine of the University of Porto, Alameda Prof. Hernâni Monteiro, 4200-319 Porto, Portugal
[d] Department of Ophthalmology, Centro Hospitalar e Universitário São João, Alameda Prof. Hernâni Monteiro, 4200-319 Porto, Portugal
[e] ESAT-PSI, KU Leuven, Kasteelpark Arenberg 10, 3001 Leuven, Belgium
[f] Department of Information Technology, University Hospitals Leuven, Herestraat 49, B-3000 Leuven, Belgium
[g] Hasselt University, Agoralaan building D, 3590 Diepenbeek, Belgium
[h] University of Antwerp, Department of Biology, 2610 Wilrijk, Belgium
[i] Flemish Institute for Technological Research (VITO), Boeretang 200, 2400 Mol, Belgium

*corresponding author
Contact details
Affiliation: KU Leuven, VITO
Postal address: Vito Health, Industriezone Vlasmeer 7, 2400 Mol, Belgium
E-mail: ruben.hemelings@kuleuven.be

Financial Support: None
Running head: Visual field estimation from OCT using deep learning
Conflict of Interest: No conflicting relationship exists for any author
Meeting presentation: the Association for Research in Vision and Ophthalmology Annual Meeting (ARVO2021)

Synopsis:
Fast, consistent visual field prediction from unsegmented optical coherence tomography scans could become a solution for visual function estimation in patients unable to perform reliable VF exams.

## Abstract

**Background/Aims:** Standard Automated Perimetry (SAP) is the gold standard to monitor visual field (VF) loss in glaucoma management, but is prone to intra-subject variability. We developed and validated a deep learning (DL) regression model that estimates pointwise and overall VF loss from unsegmented optical coherence tomography (OCT) scans.

**Methods:** Eight DL regression models were trained with various retinal imaging modalities: circumpapillary OCT at 3.5mm, 4.1mm, 4.7mm diameter, and scanning laser ophthalmoscopy (SLO) en face images to estimate mean deviation (MD) and 52 threshold values. This retrospective study used data from patients who underwent a complete glaucoma examination, including a reliable Humphrey Field Analyzer (HFA) 24-2 SITA Standard VF exam and a SPECTRALIS OCT scan using the Glaucoma Module Premium Edition.

**Results:** A total of 1378 matched OCT-VF pairs of 496 patients (863 eyes) were included for training and evaluation of the DL models. Average sample MD was -7.53dB (from -33.8dB to +2.0dB).

For 52 VF threshold values' estimation, the circumpapillary OCT scan with the largest radius (4.7mm) achieved the best performance among all individual models (Pearson r=0.77, 95% CI=[0.72-0.82]). For MD, prediction averaging of OCT-trained models (3.5mm, 4.1mm, 4.7mm) resulted in a Pearson r of 0.78 [0.73-0.83] on the validation set and comparable performance on the test set (Pearson r=0.79 [0.75-0.82]).

**Conclusion:** DL on unsegmented OCT scans accurately predicts pointwise and mean deviation of 24-2 VF in glaucoma patients. Automated VF from unsegmented OCT could be a solution for patients unable to produce reliable perimetry results.

## Introduction

Glaucoma causes retinal ganglion cell (RGC) loss, resulting in structural and functional changes in the visual system. Standard automated perimetry (SAP) is the reference technique to follow functional visual field (VF) loss during glaucoma management.[1,2] Current SAP devices such as the Humphrey® Field Analyzer (HFA) have high intra-subject variability and a lengthy examination time.[3,4] Retinal nerve fiber layer (RNFL) thickness measurements in circumpapillary optical coherence tomography (OCT) scans are standard practice for the quantitative assessment of structural retinal damage and RGC loss during glaucoma progression.[5]

The structure-function relationship in glaucoma is an intensively studied topic.[6–9] Functional damage only becomes noticeable with current SAP when extensive RGC loss has occurred.[10] SAP detects progressive glaucomatous damage better than OCT when VF loss has already occurred.[11] Studies that use OCT to estimate visual field report limited accuracy. Most models rely on assumptions such as log transformation to predict decibel (dB) VF values from retinal layer thickness values.[12–14] Deep learning (DL) approaches overcome these constraints because they can model non-linear functions.[15–19] Still, most current DL developments have limits because they (1) estimate global VF indices such as mean deviation (MD)[15,16], or (2) use OCT-derived information such as RNFL thickness maps that require a priori retinal layer segmentation[17–19].

This study goes beyond state-of-the-art by using unsegmented SPECTRALIS® OCT scans to develop and evaluate DL models that estimate the visual field sensitivity threshold at each location (52 threshold values) and MD as measured by the HFA.

## Materials and Methods

Data initially extracted comprised 1643 matched OCT-VF pairs corresponding to the 998 eyes of 542 patients who visited the University Hospitals Leuven's glaucoma clinic between 2015-2019. This work is part of the larger study on "automated glaucoma detection with deep learning" (study number S60649), approved by the Ethics Committee Research UZ/KU Leuven in November 2017. Informed consent was waived due to the retrospective nature and patient reidentification made impossible as the link between patient ID and study ID was deliberately removed. The research adhered to the tenets of the Declaration of Helsinki. Inclusion criteria were: (1) the availability of a SPECTRALIS® OCT (Heidelberg Engineering, Heidelberg, Germany) scan using the Glaucoma Module Premium Edition (GMPE), containing one scanning laser ophthalmoscopy (SLO) en face image, 24 radial scans and three circumpapillary rings, and (2) the results of an HFA3 exam with the strategy 24-2 SITA Standard (54 test points) obtained with the Humphrey® Field Analyzer (model 850 v1.3.1.2, Carl Zeiss Meditec, Dublin, CA, USA). We used the circumpapillary rings as these scans cover all nerve fibers that pass through the optic nerve, unlike the radial scans. The SLO image served as intra-study benchmark.

The 1643 OCT-VF pairs of 542 patients were allocated to train, validation, and test set, accounting for 60/20/20 of the patients, respectively. We took care that all data from a single patient were stored in the same partition to avoid leakage. Subsequently, VF data of the validation and test data were filtered on standard HFA reliability indices, with false positives (FP), false negatives (FN), and fixation losses (FL) not exceeding 15%, 33%, and 20%, respectively.[20] We did not filter unreliable visual field data in the training set but used it as data augmentation (noise).

OCT data were extracted in RAW format using Heyex v6.12.1 software (Heidelberg Engineering, Heidelberg, Germany). The binary files were subsequently processed with heyexReader v0.1.3, a Python package for reading Heyex OCT files. The three circumpapillary RNFL rings (3.5mm, 4.1mm, and 4.7mm) and SLO were extracted as lossless image files with dimensions 768x496 and 1536x1536, respectively. We obtained VF data with HFA3 that were analyzed in PeriData v3.5.7 (PeriData Software GmbH, Hürth, Germany). Pointwise sensitivity threshold values were extracted from the individual patients' printouts using an optical character recognition (OCR) tool developed for this task. OCR output was manually verified on 10% of the data, matching perfectly with actual threshold values. These values were paired with global indices such as MD that were exported as a comma-separated value text file by PeriData. Two VF test points were discarded in all analyses, as these are located on the anatomical blind spot (see Figure 1), resulting in 52 threshold values to model.

Eight DL models were trained using the 3.5mm, 4.1mm, 4.7mm circumpapillary rings and SLO images, for both MD and 52 threshold values. We compared the single models with ensemble (averaged) predictions. We selected the Xception[21] architecture pretrained on ImageNet[22]. The convolutional neural network (CNN) was followed by a global average pooling and convolution operation, avoiding any fully connected layers to minimize overfitting. The final convolution operation had either one or 52 filters, depending on the target (MD or 52 threshold values), featuring a linear activation to allow for regression. Models were trained using mean squared error (MSE) loss, optimized using Adam[23] with a starting learning rate of 1e-4. The latter was reduced to 75% of its value after ten epochs without improving the validation loss. Each epoch featured 300 training steps of batches containing four preprocessed and augmented images. Single-channel input images were either upscaled to 768x512 (circumpapillary scans) or downscaled to 512x512 (SLO), with intensity values rescaled between 0 and 1. Augmentation included horizontal flipping, elastic deformation, and cutout[24]. Model development and evaluation were performed using Keras[25] v2.2.4, Tensorflow[26] v1.12.0, in a Python 3.6.7 environment running on a server with six GTX 1080 Ti and two TITAN V graphics processing units.

The coefficient of determination ($R^2$), Pearson's r (r), and mean absolute error (MAE) were the metrics to evaluate model performances. $R^2$ and MAE were computed using the scikit-learn library[27], and r using the NumPy library[28]. The best model configuration (one for MD, one for threshold values) was selected on the highest $R^2$ metric and evaluated with the independent test set. We computed a baseline MAE for validation and test sets by predicting the mean MD or threshold value. 95% confidence intervals (CI) were obtained through bootstrap sampling (5000 iterations). The model trained on SLO images offered an objective way to assess the added value of OCT (with complete retinal layer information) over en face retinal imaging. We visualized correlation per individual VF point and VF sector (Garway-Heath[29]) to verify which VF regions are better modeled than others. Finally, we compared the 90% CI of VF sensitivity threshold predictions against empirically established test-retest variability published by Artes et al.[3]

## Results

### Study sample

Study sample characteristics are presented in Table 1. Of note, the average MD was -7.53dB (ranging from -33.8dB to +2.0dB), which is expected considering that the data come from a tertiary glaucoma clinic. After filtering on standard reliability indices, 1378 OCT-VF pairs were eligible for model development and evaluation. The baseline MAE values for MD and threshold estimation were 7.15dB and 8.17dB for the validation set and slightly lower for the test set (6.26dB and 7.64dB, respectively, see last column of Table 2).

### MD estimation

The CNN model with an Xception encoder explained up to 71% (95% CI: 0.62-0.78) of the variance in the validation set of 190 OCT-VF pairs, equivalent to an average correlation coefficient of 84%. The MAE of 3.33dB obtained using the 4.7mm scan reduced the baseline MAE by 53%. The performances of single models were similar between circumpapillary scans, but significantly outperformed the model trained with SLO images, with the latter explaining 46% (0.27-0.55) of the variance (r=0.68). The averaging of predictions of the four single models resulted in improved evaluation metrics: $R^2$=0.74 (+0.03), r=0.87 (+0.02), MAE = 3.21 dB (-0.12). This ensemble model had similar results on the independent test set of 182 OCT-VF pairs ($R^2$=0.75, r=0.87). Table 2 (second row of each cell) gives a detailed overview of MD estimation results. We display four test set cases featuring varying VF severity levels in the bottom panel of Figure 1.

### Pointwise sensitivity threshold estimation

The model trained using 4.7mm scans obtained the best results on the validation set, explaining 57% (0.48-0.63) of the variance across the 52 points, equivalent to an average correlation coefficient of 77%. The models' performances were similar ($R^2$ from 0.54 to 0.57), with $R^2$ increasing with a larger ring diameter. Similar to what we observed in the MD estimation experiments, the SLO-trained model reports significantly lower metrics ($R^2$=0.39, [0.30-0.46]). The best single-scan model (4.7mm) lowered the baseline MAE to 5.08dB, representing a reduction of 38%. For pointwise VF prediction, including the SLO-trained model in the ensembling process did not result in higher performance. Hence, the average of the predictions of the models trained using circumpapillary scans was taken, scoring $R^2$=0.59 (+0.02), r=0.78 (+0.01), and MAE=4.86 (-0.22). The test set results are similar, with $R^2$ and r at 58% and 76%, respectively.

We can detect differences in Figure 2A when inspecting the individual threshold values. The model trained on 4.7mm circle scans obtained high r values (range: 0.64-0.86) for all 52 VF test points, with the highest values recorded in superior and inferior nasal VF sectors, corresponding to nasal step locations, and the lowest in the temporal VF, corresponding to the temporal wedge location. The SLO-trained model yields lower r values (range: 0.47-0.73), especially in the inferior VF area (Figure 2B). Figure 2C illustrates this contrast, with differences up to 0.26 recorded between the two models. The best model using three OCT scans' predictions equivalently reaches high r values on all 52 test points (range: 0.68-0.87, Figure 2D). The lowest and highest values are recorded in the central and inferior nasal VF, respectively. These findings corroborate with the sectoral analysis in Table 3. The MAE baseline is the most elevated in both inferior nasal and superior nasal VF sectors. The model explains most of the variance in those sectors with R² equal to 0.64 and 0.61, respectively. The lowest metrics are in the central sector (R²=0.53, r=0.77).

Figure 3 visualizes the individual threshold prediction performance in the same box plot style as Zhu et al. (2010)[12] and Guo et al. (2017)[14]. The graph plots the SAP-measured dB values against the predicted dB values at an interval of 2dB. The largest prediction errors occurred in VF points with low sensitivity values in all three studies. However, the variability of predictions by our CNN was much more consistent with previously published test-retest CI: 33 out of 38 box plot whiskers fall within the 90% CI determined by Artes et al.[3]

## Discussion

This study is the first to regress all 24-2 VF sensitivity threshold values and MD from unsegmented OCT images, with correlations of 0.79 and 0.87 with SAP ground truth, respectively. The OCT-trained CNN models yield state-of-the-art results on VF estimation using unsegmented OCT data. Our data-driven modeling approach overcomes the need for retinal layer segmentation and prior assumptions on the structure-function relationship.

The advantages of omitting a mandatory retinal layer segmentation processing step are twofold. First, it alleviates potential segmentation errors because of bad scan quality[30,31] or critically thinned RNFL (floor effect) in severe glaucoma cases. The models allow extracting relevant information from other RNFL parameters and retinal layers besides the commonly used RNFL thickness values. Previous work hinted that OCT reflectance data might be more informative for glaucoma than conventional RNFL thickness values.[32,33] Christopher et al. verified that original voxel information from the RNFL layer resulted in CNN models with higher MD estimation performance than models trained using RNFL thickness values (0.70 and 0.63 in R² score, respectively).[15] Our SLO-trained model for MD estimation gave the same results as the Christopher et al. study (R² = 0.48 [0.41-0.54] vs. our R² = 0.46 [0.27-0.55]). Our best model (ensemble of four models trained on three types of circumpapillary rings and SLO) explains 75% (0.67-0.82) of the MD variance in the test set, whereas Christopher et al.'s best setup using average RNFL OCT voxel intensity explains 70% (0.64-0.74). Formal inter-study comparison is not possible because the evaluation metrics depend on data set characteristics such as sample MD (-7.5dB versus -5.2dB in their glaucoma subset). Another recent study by Yu et al. describes the prediction of global VF indices using a 3D CNN that takes the complete volumetric OCT scans of the ONH and macula as inputs.[16] For MD, the authors report a Pearson correlation coefficient of 0.86 (0.83-0.89), which corresponds to the correlation of 0.87 (0.83-0.91) of our ensemble model. Again, direct comparison is difficult because of different data (sample MD of -2.1dB). The high memory demands of 3D CNNs forced the authors to compromise on OCT scan resolution: the original cubes were downsized to 32% of their original size. By doing so, they introduced a risk of unintentionally removing fine-grain structural features. Our 2D CNN setup preserved the original resolution of all OCT scans (768 A-Scans). Additionally, the GMPE protocol generates the circumpapillary rings by averaging over 16 consecutive B-Scans, resulting in high-quality scans.

We report the highest correlation to date on individual threshold values of HFA 24-2 SITA Standard VF exams from OCT. Similar to MD analyses, OCT-trained models on threshold values significantly outperformed their SLO counterpart. Although not significant, the circumpapillary scans with a larger diameter seem to explain more variance in the validation data, with R² increasing from 0.54 to 0.57. This result makes sense because OCT data at 16° intersection potentially offer more information since the RGC axons are lying further apart from each other at a greater distance from the disc border. Combining the three circumpapillary scans through prediction averaging gave a correlation of five percentage points higher than the 0.74 reported by Guo et al. The latter authors obtained their results using 9-field OCT data covering 60° of the retina, whereas an area of 16° around the ONH was sufficient in our case. The three main differences between our approach and Guo et al.'s are (1) the use of CNNs versus Support Vector Machines (SVM), (2) data-driven selected image features versus thickness values of RGC complex layers, and (3) a significantly larger study sample (863 versus 86 eyes). The recent study by Park et al. describes a similar approach featuring RGC complex layer segmentation.[18] However, they employed an Inception-v3 CNN instead of an SVM to predict the 24-2 VF map. The authors did not publish metrics such as R² and correlation, hampering comparison with our results.

The ensemble model of circumpapillary scans could model specific VF points and sectors better than others. Three out of the five lowest r values were recorded in the temporal VF sector in the validation set, while the five best were all in the superior nasal VF sector. These findings match the superior nasal step scotoma location that is typically affected early in glaucoma development.[34] Lower performance in both central and temporal VF sectors could be due to damage that occurs solely in later disease stages. This result corroborates with the sectors featuring the lowest MAE baseline (Table 3), showing lower variance in ground truth threshold values. Individual r values obtained by Guo et al (based on 60° FOV data, see Figure 2E) also feature higher performance in inferior nasal and superior nasal VF sectors (up to 0.85) but significantly lower in temporal VF (as low as 0.50). Their sector average Pearson r in superior nasal and inferior nasal are close to the averages reported in this study (0.78-0.80 vs. 0.78-0.82, respectively). Besides MD, Christopher et al. predicted sectoral pattern deviation (which is correlated with threshold values) using DL approaches, equally obtaining the best performance in superior nasal VF ($R^2$=0.67) and inferior nasal VF ($R^2$=0.60).

VF testing suffers from intra-subject variability, complicating the diagnosis of glaucoma progression.[35] The most objective way of assessing VF reliability is through test-retest setups. In the Ocular Hypertension Treatment Study, VF abnormalities were not confirmed in 86% of the original reliable VF exams.[36] Average correlation between repeat HFA 24-2 exams in patients with glaucoma was 0.83 in a pilot study.[37] If this shows the theoretical maximum, then our correlation of 0.79 comes close. As Guo et al. also state, it becomes harder to assess actual performance improvements in VF modeling from OCT, given that the ground truth VF is noisy. Artes et al. computed repeat VF on 49 glaucomatous eyes, and they published 5th and 95th limits for VF threshold values.[3] Their confidence intervals provide additional evidence that VF points with lower recorded dB values hold more variability than those with higher dB values. Two studies on 24-2 VF estimation from OCT provide a comparison of measured versus predicted VF threshold values, which can be placed next to the empirical 90% CI of Artes et al. Figure 3 showcases the prediction variability for all three studies (current, Guo[14], Zhu[12]). The 90% CI in the current study show that no systematic overprediction of dB values occurred, with 33 out of 38 whiskers (88%) falling within the shaded area. This result is a significant improvement compared to the two previous non-DL studies, which were sensitive to overprediction (58% and 39% of whiskers in the shaded area). These results confirm that future performance improvement in the current model will be hard to detect, as almost all predictions fall within the empirically determined VF ranges.

This study comes with strengths and limitations. We have trained and validated our method on data sourced from the same hospital, a single type of OCT device for a single type of VF exam. We should further validate our trained models on external OCT-VF data available in the public domain. We envisage this will become possible soon, given the widespread interest in DL in glaucoma management.[38] In this study, we have taken all precautionary measures to prevent overfitting: no fully connected layers in CNN and a single use of the independent test set. Next to unknown generalizability, we provide no analysis on explainability. The move from OCT segmentation parameters to complete data-driven modeling eliminates the risk of segmentation errors but comes at the cost of model decision transparency. We plan to investigate what retinal layers are the most informative for DL-predicted VF maps in future studies.

Our findings suggest that deep learning can accurately map the structure-function relationship in glaucoma patients. Fast, consistent VF prediction from unsegmented OCT could become a solution for visual function estimation in patients that cannot perform reliable VF exams.

Acknowledgments

The Research Group Ophthalmology, KU Leuven and VITO NV jointly supported the first author. This research received funding from the Flemish Government under the "Onderzoeksprogramma Artificiële Intelligentie (AI) Vlaanderen" programme. No outside entities have been involved in the study design, in the collection, analysis, and interpretation of data, in the writing of the manuscript, nor in the decision to submit the manuscript for publication. Thus, the authors declare that there are no conflicts of interest in this work.

## Images

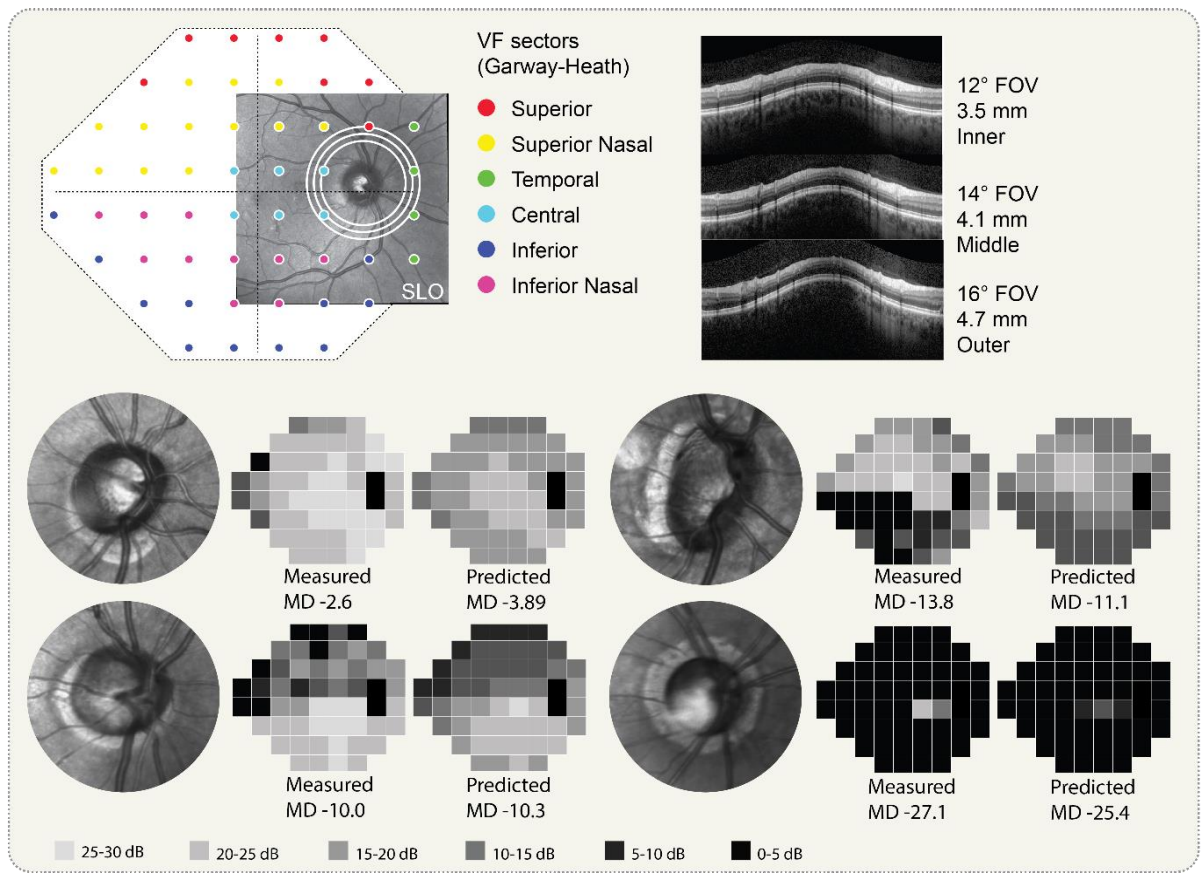

Figure 1 – (top panel) Overview of imaging modalities, the spatial relationship between structure and analyzed VF points, allocated to Garway-Heath sectors. The 30° SLO covers less than half of VF threshold values, still considerable more than the three circumpapillary OCT scans (white circles) displayed on the right. (bottom panel) Four cases of the independent test set. Each case features (1 ) an ONH-zoom of the original 30° SLO image, (2) measured VF map and MD, and (3) the corresponding predicted VF map and MD. The displayed cases include an example of early glaucoma (top left), moderate glaucoma with loss in superior hemifield (bottom left), a myopic eye with severe glaucomatous loss in inferior hemifield (top right), and severe glaucoma with only a small central island left (bottom right).

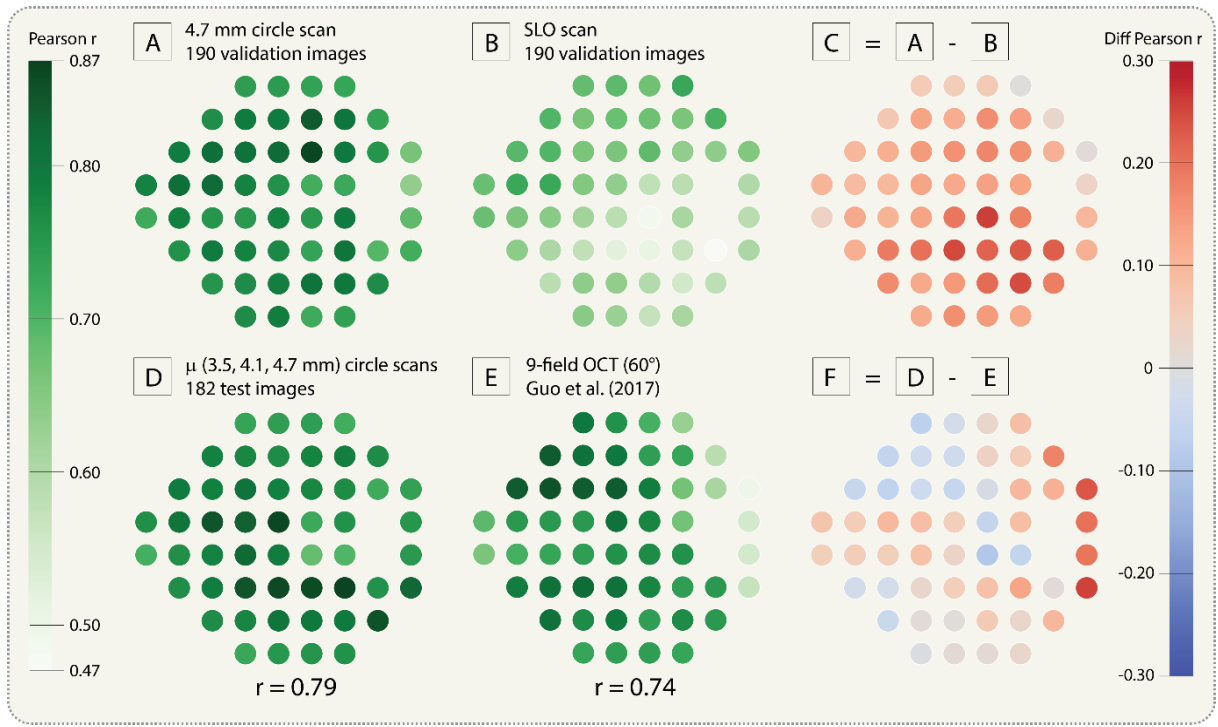

Figure 2 – [A] Pearson r values for 52 VF threshold values obtained using model trained on 4.7mm (outer) OCT scan. [B] Similar to [A], but model trained using en face SLO images. [C] visualizes the difference between [A] and [B], indicating the superior performance of OCT scans in the inferior hemifield. [D] Final Pearson r values obtained on the test set, using the averaged predictions of the three OCT scans (r=0.79). [E] Best results published prior to our work, obtaining a global r of 0.74.[14] [F] visualizes the differences between the results presented in this study and previous state-of-the-art.

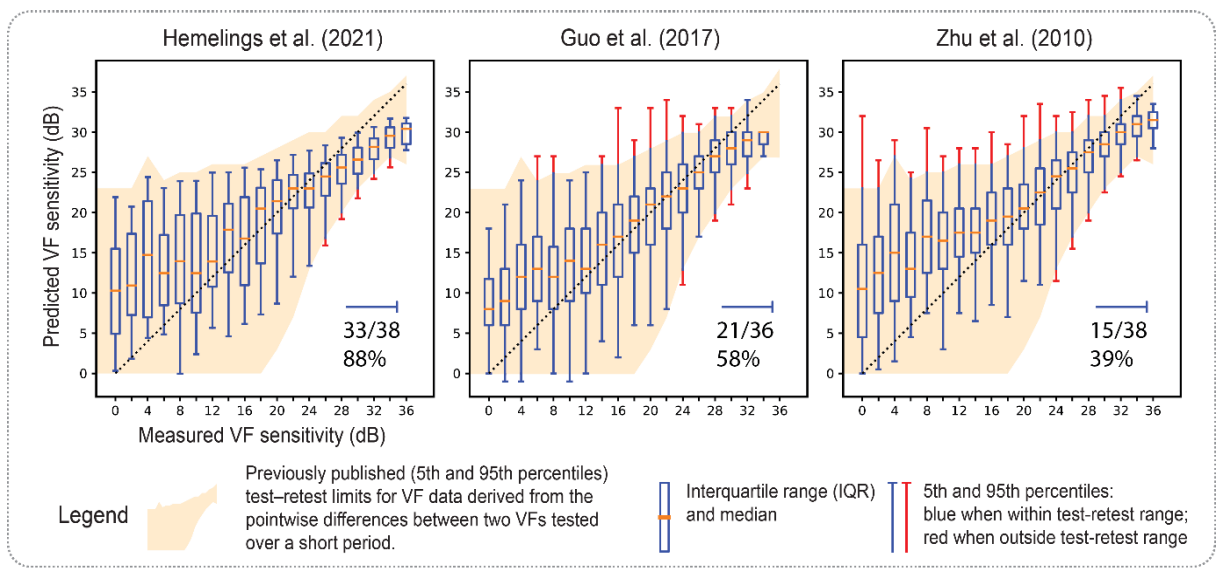

Figure 3 – Comparative overview of three original studies (current, Guo[14], Zhu[12]) that report on the relationship between measured and predicted VF threshold values, stratified by sensitivity (step size of two dB). The error ranges obtained by our approach leveraging DL are smaller than previous non-DL studies. 33 out of 38 whiskers are located within the 90% CI test-retest limits reported by Artes et al.[3]

# Tables

Table 1 – Study sample characteristics

**Table 1 – Study sample**

|  | Train | Val | Test | Total |
|---|---|---|---|---|
| OCT-VF pairs | 1006 | 190 | 182 | 1378 |
| Eyes | 598 | 135 | 130 | 863 |
| Patients | 325 | 83 | 88 | 496 |
|  |  |  |  |  |
| Age | 55.8 ± 19 | | | |
| Sex (F \| M) | 0.50 \| 0.50 | | | |
| MD data available | 1378/1378 (100%) | | | |
| MD (dB) | -7.53 ± 7.77 | | | |
| MD ≥ -6 dB | 829 (60%) | | | |
| -6 dB > MD > -12 dB | 233 (17%) | | | |
| MD ≤ -12 dB | 316 (23%) | | | |
| SphEq data available | 1091/1378 (79%) | | | |
| SphEq (D) | -2.28 ± 2.78 | | | |
| +1 D ≤ SphEq (hyperopia) | 116 (11%) | | | |
| +1 D > SphEq > -1 D (emmetropia) | 237 (22%) | | | |
| -1 D ≥ SphEq > -6 D (myopia) | 617 (57%) | | | |
| -6 D ≥ SphEq (high myopia) | 121 (11%) | | | |

Table 2 – Quantitative results for all models trained for the estimation of 52 threshold values (first row of each cell) and MD (second row of each cell). The first section features results on the validation set (190 images), for which the best results are set in **bold**. Best model setup on validation data was subsequently used to obtain results on the independent test set (182 images). MAE baseline for validation and test data was computed through the constant prediction of the mean value (threshold value, MD).

| Modality | Target | R² [95% CI] | Pearson r [95% CI] | MAE (dB) [95% CI] |
| --- | --- | --- | --- | --- |
| Baseline (validation) | Threshold values | 0.00 | 0.00 | 8.17 |
|  | MD | 0.00 | 0.00 | 7.15 |
| Inner \| 3.5mm | Threshold values | 0.54 [0.45-0.61] | 0.75 [0.69-0.80] | 4.97 [4.49-5.45] |
|  | MD | 0.71 [0.62-0.78] | 0.84 [0.79-0.88] | 3.40 [2.93-3.89] |
| Middle \| 4.1mm | Threshold values | 0.55 [0.45-0.62] | 0.76 [0.70-0.80] | 5.06 [4.62-5.53] |
|  | MD | 0.67 [0.53-0.77] | 0.83 [0.77-0.88] | 3.65 [3.06-4.05] |
| Outer \| 4.7mm | Threshold values | 0.57 [0.48-0.63] | 0.77 [0.72-0.82] | 5.08 [4.68-5.50] |
|  | MD | 0.71 [0.60-0.80] | 0.84 [0.78-0.90] | 3.33 [2.71-3.67] |
| SLO | Threshold values | 0.39 [0.30-0.46] | 0.65 [0.58-0.71] | 5.79 [5.23-6.38] |
|  | MD | 0.46 [0.27-0.55] | 0.68 [0.54-0.76] | 4.82 [4.18-5.44] |
| Circle scans (avg) | Threshold values | **0.59 [0.50-0.65]** | 0.78 [0.73-0.83] | 4.86 [4.43-5.30] |
|  | MD | 0.73 [0.64-0.82] | 0.86 [0.81-0.91] | 3.19 [2.62-3.52] |
| Circle scans, SLO (avg) | Threshold values | 0.59 [0.51-0.64] | 0.79 [0.74-0.83] | 4.88 [4.44-5.33] |
|  | MD | **0.74 [0.66-0.81]** | 0.87 [0.82-0.91] | 3.21 [2.67-3.53] |
| Baseline (test) | Threshold values | 0.00 | 0.00 | 7.64 |
|  | MD | 0.00 | 0.00 | 6.26 |
| Test set (182 images) | Threshold values | 0.58 [0.51-0.64] | 0.79 [0.75-0.82] | 4.76 [4.40-5.14] |
|  | MD | 0.75 [0.67-0.82] | 0.87 [0.83-0.91] | 2.84 [2.47-3.24] |

Table 3 – Metrics on the six visual field sectors as described by Garway-Heath[29], computed on the test set. MAE baseline was obtained by always predicting the sector threshold mean. The most relevant VF sectors for glaucoma are highlighted in bold.

| Sector | R² | Pearson r | MAE (dB) | MAE baseline |
| --- | --- | --- | --- | --- |
| Central | 0.53 [0.46-0.58] | 0.77 [0.72-0.81] | 4.67 [4.17-5.19] | 7.35 |
| Temporal | 0.55 [0.44-0.63] | 0.77 [0.70-0.83] | 4.33 [3.94-4.74] | 6.41 |
| Inferior | 0.56 [0.46-0.64] | 0.77 [0.70-0.82] | 5.05 [4.62-5.52] | 8.05 |
| **Inferior Nasal** | **0.64 [0.55-0.70]** | **0.82 [0.77-0.86]** | **4.64 [4.16-5.11]** | **8.22** |
| Superior | 0.54 [0.44-0.62] | 0.76 [0.70-0.81] | 4.89 [4.47-5.31] | 7.43 |
| **Superior Nasal** | **0.61 [0.52-0.68]** | **0.80 [0.75-0.84]** | **4.71 [4.24-5.21]** | **8.18** |